\documentclass[journal]{IEEEtran}

\usepackage{graphicx,amsmath,amssymb}

\usepackage{multicol,multirow} 
\usepackage{url} 

\usepackage{appendix,titlesec}  
\titleformat{\section}{\normalfont\Large\bfseries}{\thesection}{1em}{}  

\usepackage{array}
\newcolumntype{C}[1]{>{\centering\arraybackslash}m{#1}}

\usepackage{acronym}

\begin{document}
\title{Data Requirements and Prediction Scaling for\\Long-Term Failure Forecasts in Wind Turbines 
}

\author{Viktor Begun and Ulrich Schlickewei\\
Technische Hochschule Ingolstadt, Esplanade 10, D-85049 Ingolstadt, Germany
}

\maketitle

\begin{abstract}
We investigate the key factors that enable early failure forecasting in wind turbines.
For this purpose, we analyze studies with long-term forecasts and compare their main features: prediction time, methods, targeted components, dataset size, and check the effect of using additional sensors. 
We found that the size of the dataset is the main factor and that an approximate linear scaling holds: the number of forecast days is twice the size of the dataset, measured in turbine years. 
We also observe that the data allow us to quantify the meaning of "big" and "long" in the terms "big data" and "long-term" forecasts, which are found to be ten turbine years and two weeks. 
\end{abstract}


\IEEEpeerreviewmaketitle

\section{Introduction}
\IEEEPARstart{T}{his} study aims to systematize approaches to long-term failure forecasting in wind turbines and identify the key factors that enable such predictions. 
We start defining long-term forecasts as predictions made at least two days in advance -- a requirement commonly observed among wind turbine operators, see, e.g., the wind energy challenge raised by Energias de Portugal~\cite{EDP2018}. 
This requirement aligns with the fact that powerful wind turbines are large and complex structures with high repair and downtime costs. Their maintenance requires skilled personnel and specialized equipment, making it necessary to allocate time for deciding which repair team to deploy and finding and delivering the required spare parts. 
This minimal required prediction time may further increase due to the ongoing trend towards larger and more powerful turbines~\cite{wiser2021expert}.  

The size of the available dataset may significantly influence the analysis approach and its outcome. Some methods are better suited for small datasets, while others perform more effectively with large datasets. Additionally, a minimal dataset size may be required for meaningful analysis. 
Since obtaining large datasets is often challenging~\cite{Kusiak2016}, it is essential to understand how dataset size impacts prediction time and fault detection methods. 
This knowledge enables practitioners to estimate the required dataset size and choose appropriate analysis methods to predict failures with the desired prediction horizon. 

We propose measuring dataset size in turbine years (TY), a standardized data unit representing the data collected from a single turbine over one year. 
The exact amount of data in a TY depends on the number of sensors and the recording interval, but is generally millions of data points. 
For example, assuming that the data is obtained by the standard Supervisory Control and Data Acquisition (SCADA) system with 10-minute intervals (144 recordings per day), a turbine has 10 key components, each monitored by 4 measured values (e.g., minimum, maximum, average and standard deviation during the 10-minute interval), and one year comprising approximately 365.25 days, this totals 10*4*144*365.25~$\simeq$~2.1 million data points per TY. 
The exact number of components and the measured values may vary depending on the turbine type, and the recording interval may be shorter, particularly if additional vibration sensors are used.

However, we argue that a time-based measure like TY is better suited for a performance estimate than a data-volume-based measure, which requires precise counting of data points. 
The reason is that we analyze commercial wind turbines currently available on the market. Due to competition between manufacturers, these turbines are optimized for the key performance parameters most important to customers. Consequently, it is not the volume of the data flow between turbines and analysis systems that matters, but rather the years a turbine operates profitably -- in other words, turbine years. 
The term TY is similar to labor hour, machine hour, and horsepower, and its purpose is the same -- standardization and the possibility to compare. The term TY exists and is used, for example, to compare the size of datasets~\cite{dao2019wind} or to quantify the harm of wind turbines to birds~\cite{mcclure2021eagle,ferrer2022significant}, but we see a margin for increasing the popularity of this term.
Furthermore, our interest lies in a rough estimate that distinguishes small datasets from a few turbines and large datasets comprising hundreds of turbines in a wind park, which TY reflects effectively.  

In addition, we would like to find out how much and for which cases additional sensors can improve the forecast horizon. The reasons are that standard SCADA sensors are cheaper, they are already installed in thousands of turbines, which are often situated in remote places, the corresponding maintenance protocols are developed and optimized, and maintenance teams are trained. The profit from additional sensors should overcome the burden of installing and maintaining them, while the profit is obviously related to the advantage in the prediction time of failures.

We found that our threshold to predict failures at least two days in advance pass only the studies that are using the datasets of the size larger than 0,4~TY. However, we observe that the data-driven scales are 10~TY for the dataset size and 2 weeks for the duration of a prediction. 
Ten turbine years is an approximate separation scale when analysis methods change from advanced specifically designed methods and usage of additional vibration sensors for small datasets to basic statistical techniques and machine learning for larger datasets. 
Two weeks is the lowest prediction horizon, which appears, when researchers have datasets larder than 10~TY. 
Therefore, ten turbine years and two weeks can be taken as data-driven definitions for a dataset to be called big and a forecast to be called long-term. 
We also observed a positive correlation between the size of the dataset and the prediction time, with an approximate scaling \texttt{prediction days}\,$\sim~2*$TY. 
Data from vibrational sensors allow for order of magnitude better prediction horizons than those suggested by linear scaling for small datasets. 
However, big datasets and dedicated specific approaches allow one to obtain similarly good predictions for small datasets using standard SCADA data.  

The paper is organized as follows. Section~\ref{Sec:Methodology} describes the methods we used to find relevant papers. Section~\ref{Sec:Results} shows our results and findings, while Section~\ref{Sec:Conclusions} concludes the paper.

%
\section{Methodology}\label{Sec:Methodology}
%

The number of research articles on wind turbines is enormous and growing rapidly. A search for ”wind turbine” in Scopus in July 2024 gave us more than 100,000 indexed papers. Moreover, this number grows with acceleration. There were 5,000 papers indexed only in one year of 2013, while in 2023 this number of indexed papers was about 9,000. Limiting the search results only to the words ”wind” and ”turbine” found one after another with zero words in between (wind w/0 turbine), only to articles published in journals (no books and no proceedings), and only in the English language reduces the mentioned numbers to about 47,000, 1,600 and 4,800 correspondingly, with the most important factor being the proceedings exclusion. This abundance means that it is practically impossible even to read every article published within five recent years, and one must rely on some filtering. Therefore, we set our aim not to cover every paper but rather show popular solutions and discover the main underlying methods and trends.

Our approach is twofold. On the one hand, we collected the papers from the literature that were relevant for a research project on fault prediction of wind turbines in the context of a specific case study with an open dataset, see Ref.~\cite{begun2024cost}. These papers we found using Google Scholar. 
On the other hand, we used Scopus to make a thorough scan of articles published within the last five years that we found for a search phrase "wind w/0 turbine and SCADA and \texttt{signal}", where \texttt{signal} should be replaced with a term meaning a specific signal, for example, vibration. 

The use of Google Scholar is motivated by the fact that it is free of charge, counts citations well, thus showing approaches that are relevant to others, and is widely available outside academic institutions. 
The use of Scopus is motivated by the fact that it is a curated proprietary database that allows complicated searches, counts citations as well, and sorts by year even better. 

The choice of SCADA and a \texttt{signal} was to limit the number of papers for deep reading and to check a popular hypothesis that modern techniques allow making good predictions using only SCADA data without dedicated and typically expensive additional sensors. The \texttt{signal} measures that we used in the Scopus search include the mentioned "vibration" and also "strain", "torque", "LiDAR", and "acoustic". We also tried to search other signals, see Ref.~\cite{badihi_comprehensive_2022} for a comprehensive list of possibilities. However, they give either too rare results in combination with SCADA, e.g. "thermography", or too many false positives as "electric".   

The most common is SCADA data with a 10-minute acquisition frequency. In principle, 5-, 1-minute, and other SCADA frequencies exist, but they are rare. The change from 10- to 1-minute frequency increases the amount of data 10 times but still does not allow monitoring fast processes. 
A radically different amount of data appears if one considers vibrations because even a 1-minute frequency is only 1/(60\,s) $\simeq$ 0.017~Hz, while vibration sensors measure values of the order of kHz. 
Therefore, we do not multiply the amount of more frequent SCADA data by a corresponding factor, but treat the SCADA data, for which vibration measurements are available, separately.

A turbine consists of many interrelated components that form different groups. To classify failures into a small number of categories, we generalize compound terms based on our understanding of which element of the compound name better describes the failure. 
For example, the name of the \texttt{generator cooling} failure discussed in Ref.~\cite{borchersen2016model} consists of two words. We choose to omit the word \texttt{cooling} and leave the word \texttt{generator} because it addresses the overheating of the generator, which is one of the most expensive components. This choice may not be optimal in all cases, because one may be interested also in all possible cooling failures. Another example is \texttt{generator bearing} and \texttt{gearbox bearing} failures discussed in Ref.~\cite{schlechtingen2011comparative}, which we classify as \texttt{bearing} failures because we think that these failures are rooted in bearings. 

We analyze only the best predictions of the authors and disregard false alarms, failures that have not been found, absence of failure decision mechanism, validation or test, or both, and also accept the prediction if authors at least see a failure with their approach. The latter is particularly common for small datasets with TY$<$10. 
If several predictions are available, then we take an arithmetic mean. 
These are subjective choices. However, we provide citations to the discussed papers, and a reader can make his own classifications of the papers with long-term predictions that we found.

%
\section{Results}\label{Sec:Results}
\begin{figure*}[th!]
  \centering
  \includegraphics[width=0.7\textwidth]{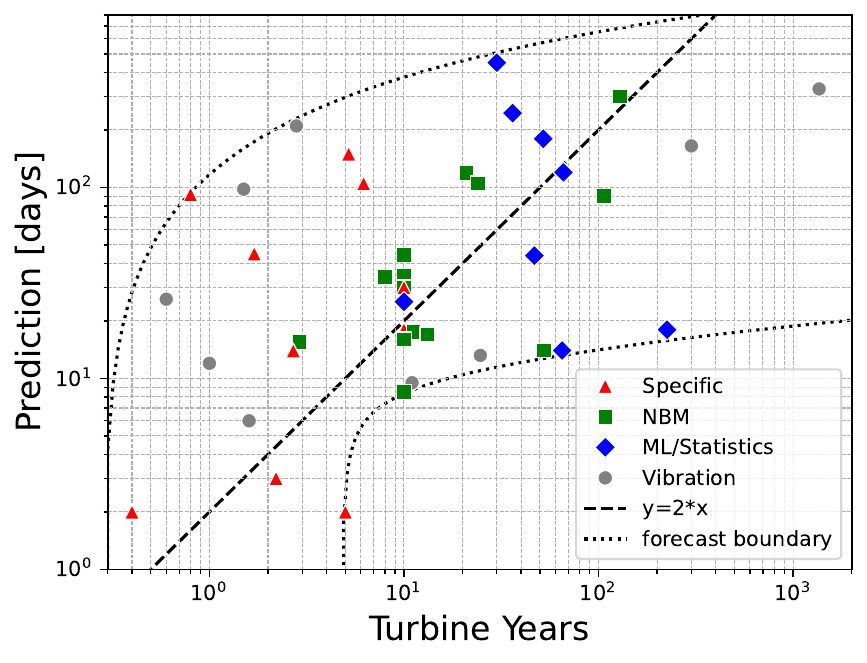}
  \caption{Prediction time as a function of dataset size measured in turbine years. The dotted lines indicate the forecast boundaries assuming saturation, while the dashed line shows linear estimation -- prediction days equal the number of turbine years multiplied by two. Labels denote four groups: predictions for small datasets with TY$<$10 (Specific), large datasets with TY$>10$ (ML/Statistic), normal behavior models (NBM), and datasets with both SCADA and vibration measures (Vibration).}\label{fig:Prediction_vs_TY}
\end{figure*}

%
%
The main result of this work is Figure~\ref{fig:Prediction_vs_TY}, 
which shows the relation between the size of a dataset and the prediction time. The selected articles are shown in Tables~\ref{Tab:TY} and~\ref{Tab:TY-Vibration}, with a corresponding list of acronyms in the Appendix. 
\begin{table*}[h!]
\centering
\caption{First author of a paper, failed component, method used for a solution, size of used dataset in TY, prediction in the number of days. The list of acronyms is in Appendix. The table is sorted according to the prediction time. The number of prediction days and TY larger than one are rounded to integers.}
\begin{tabular}{||C{3cm}|C{4cm}|C{4.5cm}|C{0.65cm}|C{0.65cm}||}
\hline
Author & Component & Method & TY & Days \\
\hline
\multicolumn{5}{||c||}{TY $<$ 10, specific methods often used} \\
\hline
Pei~\cite{pei2018wind}       &	any anomaly &	FV, MR, kNN &   5	&   150     \\
\hline
Cui~\cite{cui2018anomaly}	      & Gearbox	&   NARX, WT, NN	&  6	&   105 \\
\hline
Hu~\cite{hu2018prediction}	      &   Bearing	&  WP, MA	&    0,8	&   92  \\
\hline
Ferguson~\cite{ferguson2019standardisation}  &   Gearbox	&    Phys	&  2	&   45  \\
\hline
Garan~\cite{garan_data-centric_2022}	  &   Bearing, Gearbox	&  FI, HCF, MI, PCA, ICA, DT	& 10	&  30  \\
\hline
Miele~\cite{Miele_2022}	  &   Gearbox, Generator, Bearings, Transformer	& GCA-MTS, LSTM-AE, CNN–LSTM, GMIn, LRIn	& 10	&  26  \\
\hline
Bonacina~\cite{Bonacina_2022}  &   Bearing	& PPS	&   10	&  18  \\
\hline
L{\'a}zaro~\cite{lazaro2020determining}	  &   Gearbox	&  GMC, HIn	&  3	&   14  \\
\hline
Wang~\cite{wang2019early}	  &   Bearing	&    RBM, MD, kMC	&  2	&   3   \\
\hline
Chen~\cite{chen2018generalized}	  &   Bearing	&    CombP, IF, SVM, RBFNN	& 5	&   2   \\
\hline
Zhang~\cite{zhang2022dynamic}	  &   Gearbox, Converter, Rotor	& GRU, dAE, SVR, MD, dThr	&   0,4	&   2 \\
\hline
\hline
\multicolumn{5}{||c||}{Normal behavior model} \\
\hline
Borchersen~\cite{borchersen2016model}	 &    Generator	& KF, CUSUM	& 129	& 300   \\
\hline
Schlechtingen~\cite{schlechtingen2011comparative}	& Bearing, Stator	&   AR, NN	&    21	& 119  \\
\hline
Encalada-D{\'a}vila~\cite{encalada2021wind}	& Bearing	&   NN, BayR	&  24	& 105  \\
\hline
Tutiv{\'e}n~\cite{tutiven2022early}	    &  Bearing	&   SVM, EWMA	& 107	& 90  \\
\hline
Begun~\cite{begun2024cost}	  &   Hydraulic	& LinR, DI-CUSUM	&  10	& 44 \\
\hline
Jankauskas~\cite{Jankauskas_2023}	&    Gearbox	&   GRU, LSTM	& 10	& 35 \\
\hline
Rashid~\cite{rashid2020fault}	&    Gearbox	&   BagR	&  8	& 34    \\
\hline
Barber~\cite{barber_enabling_2022}	&    Hydraulic, Generator, Bearing	& LoMST, WHC-LOF, CUSUM	& 10	& 30   \\
\hline
Yuan~\cite{yuan2019gearbox}	&  Gearbox	&   DBSCAN, RF, GBDT, XGBOOST, MD, QF	& 11	& 18 \\
\hline
Kong~\cite{kong2020condition}	&  Gearbox	&   PPMC, CNN, GRU, EWMA	&  13	& 17   \\
\hline
Wang~\cite{wang_operating_2022}	&  Gearbox	&   DBN, DDPG-SOM	& 10	& 16\\
\hline
Yan~\cite{yan2014condition}	&   Bearing	&   NN	&    3	& 16\\
\hline
Schlechtingen~\cite{schlechtingen2014wind}	& Hydraulic, Gearbox, Converter, Anemometer, Power	&  ANFIS, FL	& 53	& 14\\
\hline
Liu~\cite{liu_condition_2022}	&   Generator, Hydraulic, Transformer	& MR, LASSO, LARS	&   10	& 9 \\
\hline
\hline
\multicolumn{5}{||c||}{TY $>$ 10, machine learning and basic statistical methods often used} \\
\hline
McKinnon~\cite{mckinnon2021investigation}    &	Pitch     &	IF    &	30  &	450   \\
\hline
Herp~\cite{herp2019assessment}        &	Bearing   &	WD, RNN, LSTM &	36  &	245   \\
\hline
Zaher~\cite{zaher2009online}       &	Gearbox   &	NN    &	52  &	180   \\
\hline
Letzgus~\cite{letzgus2020change}     &	Gearbox, Generator, Pitch, Hydraulic, Electrical  &	KCPD  &   66   &	120    \\
\hline
Zhao~\cite{zhao2016fault}        &	Generator &	PCA, DBSCAN, SVM, ARIMA   &	47  &	44    \\
\hline
Eriksson~\cite{eriksson2020machine}    &	Gearbox, Generator, Bearing, Transformer, Hydraulic   & "Kafka": NN, PCA, LinR, DT, RF, L-SVC &	10  &	25  \\
\hline
Zhao~\cite{zhao2017fault}        &	Generator &	PCA, SVM, SMOTE, DBSCAN, AOIn, SVM, NN, kNN, BayC &	225 &	18    \\
\hline
Chen~\cite{chen2013wind}        &	Pitch     &	ANFIS &	65  &	14    \\
\hline
Djeziri~\cite{djeziri2018hybrid} & general approach & PCA, BGraph, DegrV & -- &  3 \\ 
\hline
\end{tabular}
\label{Tab:TY}
\end{table*}
\begin{table*}[h!]
\centering
\caption{The same as table~\ref{Tab:TY} for the case of using additional vibration sensors.}
\begin{tabular}{||C{3cm}|C{3.5cm}|C{5.5cm}|C{0.65cm}|C{0.65cm}||}
\hline
Author & Component & Method & TY & Days \\
\hline
\hline
\multicolumn{5}{||c||}{Vibration sensors used} \\
\hline
Rezamand~\cite{rezamand2020integrated}    &	Bearing   &	WT, WA    &	1360    &   	328    \\
\hline
Natili~\cite{natili2021multi}      &	Bearing   &	SVR, PCA  &	3  &  210  \\
\hline
Carroll~\cite{carroll2019wind}     &	Gearbox   &	 SVM, LogR, NN    &	 300   &   	165   \\
\hline
Turnbull~\cite{turnbull2021combining}    &	Gearbox, Generator    &	SVM   &	 2   &   	98    \\
\hline
Teng~\cite{teng2016prognosis}	    &   Bearing    &	NN    &	0,6   &  26   \\
\hline
Zhu~\cite{zhu2021condition}         &	Gearbox   &	DAE, PCA, SVM, kNN, LOF, IF   &	 25  &   	13  \\
\hline
Cheng~\cite{cheng2018enhanced}       &	Bearing   &	PF, ANFIS &	1   &   	12 \\
\hline
Xu~\cite{xu2021fault}          &	Bearing   &	dAE   &	11    &   	10   \\
\hline
Hu~\cite{hu2022health}          &	Generator &	MA, QF    &	2   &   	6  \\
\hline
\end{tabular}
\label{Tab:TY-Vibration}
\end{table*}
We found 43 articles that give forecasts for 2 days and more:
\begin{itemize}
    \item $\geq 2$ and $<10$ days -- 6 articles~\cite{wang2019early, chen2018generalized, zhang2022dynamic, liu_condition_2022, djeziri2018hybrid, hu2022health},
    \item $\geq 10$ and $<100$ days -- 25 articles~\cite{
    begun2024cost, hu2018prediction, ferguson2019standardisation, garan_data-centric_2022, Miele_2022,
    Bonacina_2022, lazaro2020determining, tutiven2022early, Jankauskas_2023, 
    rashid2020fault, barber_enabling_2022,  
    yuan2019gearbox, kong2020condition, wang_operating_2022, 
    yan2014condition, schlechtingen2014wind, zhao2016fault, 
    eriksson2020machine, zhao2017fault, chen2013wind, turnbull2021combining, teng2016prognosis,
    zhu2021condition, cheng2018enhanced, xu2021fault},
    \item $\geq 100$ days -- 12 Articles~\cite{borchersen2016model, schlechtingen2011comparative, pei2018wind, cui2018anomaly, encalada2021wind, mckinnon2021investigation, herp2019assessment, zaher2009online, letzgus2020change, rezamand2020integrated, natili2021multi, carroll2019wind}.
\end{itemize} 

The smallest dataset in our selection is 0,4~TY, which means that one needs at least this amount of data to make a prediction for two days. 
Figure~\ref{fig:Prediction_vs_TY} shows a positive correlation between the size of the dataset and the prediction time. 
The dotted lines limit the area filled with the analyzed points, assuming the rising and then saturating dependence of the prediction time on the amount of data. These lines are obtained by enclosing the filled area using the function $a*\ln(\text{TY}-b)+c$. The parameters for the upper line are $a=120$, $b=0.15$, $c=100$ and $a=2$, $b=-4.7$, $c=5$ for the lower line.  
The dashed straight line shows a linear estimate $a*$TY with $a=2$. The straight line does not follow the form of the filled area as good as the dotted lines, but about half of the points deviate by a factor less than three from it.   
Thus, the scaling is approximately linear. 
Having the dataset of the size 0,4~$\lesssim$~TY~$\lesssim$~100, one may expect a forecast period of 2*TY days. 
Although larger datasets are expected to enable better prediction and the dependence may be nonlinear, it is useful to have a simple rule to estimate the amount of data needed for a desirable prediction horizon. 

NBM is the group of methods that best follows the linear scaling. We assign a point to the group if the authors build a model of normal behavior and then analyze the accumulated deviations from it. 
The group that does not follow the scaling and gives much better results for small datasets is Vibration. The reason is that vibration gives access to fundamentally different information with much better time resolution than SCADA. Another reason is that we don't take into account the size of the vibration dataset, which is hard to estimate because authors often do not specify vibration frequencies and the time cuts they applied to reduce the amount of vibration data. 

The lower right corner of Fig.~\ref{fig:Prediction_vs_TY} has an empty region with the inflection point around the intersection of 10~TY and the prediction time of 10-20 days.  
The existence of this region can be qualitatively interpreted so that, having large data, one is not interested in short predictions. The borders of this region quantify the meaning of \texttt{large} data, as well as the \texttt{long}-term forecast. 

The vertical boundary at about 5-10~TY can be attributed to the transition from \texttt{small} to \texttt{large} data. It also separates the Specific and the ML/Statistics groups in our notation. The distinction is not strict, as the authors of the papers that we select in these groups sometimes use the same or similar methods, which may also overlap with NBM. However, we see a tendency to use advanced mathematical methods and sophisticated modeling for datasets with TY$<$10, while for larger datasets, authors more often use "out of the box" machine learning methods for classification and have enough data to apply filters.  
The examples are the usage of the Wiener process~\cite{hu2018prediction} and the Gaussian mixture copula model~\cite{lazaro2020determining} for small datasets, while for large datasets, typical methods include direct classification using the isolation forest~\cite{mckinnon2021investigation} or a perceptron~\cite{zaher2009online} to differentiate between normal and abnormal states. Small datasets are also more often analyzed with special features designed by authors only for these datasets and not found in other works. These features are also called scores, factors, indexes, or indicators, see the exact author's namings in the list of acronyms in the appendix.  
One may notice that the list of acronyms is larger than the number of papers, and only a few acronyms repeat. This means that comparing the performance of different methods is difficult because there is not enough statistics for the methods.

The lower horizontal boundary appears at 10-20 days when the dataset size reaches 10~TY. 
There is no technical limitation preventing short-term predictions with large datasets, while 10-20 days approximately correspond to 14 days, which is two weeks. 
The border and its value may be an indication of a human-related preference. 
This preference can be interpreted so that the amount of data becomes large enough to start the prediction not from the smallest, but from some desired time horizon.  
Dataset owners may wish to achieve the longest horizons possible and demand the predictions for at least two weeks. 
Below 10~TY such a prediction is not possible, while for larger datasets, a two-week prediction de facto becomes the standard in the wind turbine industry and can be used as a data-driven definition for \texttt{long}-term forecasts.   

The upper horizontal boundary separates the empty left corner of Fig.~\ref{fig:Prediction_vs_TY}, highlighting the intuitively clear fact that long-term predictions are difficult with small datasets. 
The growth is monotonous and saturates at about 300 days, which can be interpreted so that a forecast of about a year is enough. 
Note that the lower horizontal boundary for predictions with datasets larger than 10~TY is also slowly growing. This growth may indicate that, even with enough data, it is hard to forecast for two weeks, but it becomes easier when the size of the available dataset grows.

A large number of predictions for datasets of 10~TY indicates a freely available dataset that has been analyzed by many authors, including us. 
These predictions total nine with an average forecast horizon of 26 days and a standard deviation of 11 days. This aligns well with the linear scaling discussed earlier, suggesting the possibility of a 20-day forecast for a dataset of 10~TY. However, it is important to note that the owners of the corresponding dataset imposed several constraints, including a strict upper limit: a forecast had to be no longer than 60 days, see Ref.~\cite{begun2024cost} for details. 

Excluding the unavailable regions and the anomaly at 10~TY, one can see that the points are distributed relatively homogeneously on the double-log scale. This means that the probability of having large datasets or long-term predictions drops exponentially. In other words, it is very hard to find large datasets and make long-term predictions. 

It would be good to have more data points in Fig.~~\ref{fig:Prediction_vs_TY}. However, it is difficult to find more papers due to the variety of possible methods and their combinations. It is also difficult to find a good searchable criterion. For example, one cannot search by prediction time because authors do not report it in a unified format. The size of the dataset one has to calculate from the authors' notice of the start and end of observations, which is also not unified or even not mentioned, and one has to calculate it from their figures. 
Our systematic Scopus search for "wind w/0 turbine and SCADA and vibration" brought us mainly the articles that we already have seen using Google Scholar, with the exception of Ref.~\cite{tutiven2022early}. 
The Scopus search for other sensors brought us only false positives. 
We found that the keywords strain and torque are used to examine the main structural elements such as foundation and tower of the turbine. These elements are designed for much longer services than other components of a turbine. The typical time scale is 20 years, and there is also a large safety margin. Therefore, the corresponding fatigue life evaluations can predict 20, as well as 200 years remaining life~\cite{mai2019prediction, zhao2023fatigue}. 
The "LiDAR" as a searched signal gives articles with predictions no longer than 5 minutes, but clear clustering by amount of TY. Among the 21 articles found, half analyze datasets with less than 1~TY and discuss mainly yaw and alignment, while the rest discuss mainly power reduction due to wake and aging. 
%
%
The "acoustic" gives articles with only instant measures without prediction and for typically very small datasets starting from 0.026~TY. This particular number of TY originates from 4 turbines and 2 hours of measurements every night in February~\cite{iannace2019wind}. The predicted value is not the condition of the turbine, but the noise it makes.  

Answering the question of whether SCADA and a dedicated sensor are always better than SCADA is cumbersome. 
First of all, there are no openly available datasets with SCADA and vibration data. Therefore, one has only a solution given by the analyzers of the dataset, which is not possible to check independently. Most authors did not ask themselves the question whether one can get the same prediction with SCADA only and were just showing the result of the analysis. There is also a known effect of over-presenting positive results and under-publishing negative results. 
%
%
However, we found one paper, which shows that both SCADA and vibration data give comparable results~\cite{natili2021multi}. 
Our Figure~\ref{fig:Prediction_vs_TY} also suggests that having vibration data can be beneficial, but with enough data and some effort, one can get similarly good results using only SCADA.
Note that despite the fact that vibration sensors can significantly improve prediction, vibrations are typically measured for parts with rotating elements, such as gearbox, generator, and bearing, and for the elements that are subject to an external force, such as tower, foundation, and blades, which represent a significant amount but not all turbine components. 

One can advocate for the usage of the rated power and production year with arguments similar to those for the use of TY. 
Turbines of similar power should have similar construction and hence similar failure modes and rates. 
There could also be a dependence on the calendar year in which a dataset was analyzed. 
Rapid growth in computing power, the development of machine learning algorithms, and the growing interest in condition monitoring may also be important factors. A fashion for using certain models may exist and change with time. 
The authors of Refs.~\cite{Montero_Jimenez_2020, Chatterjee_Dethlefs_2021} observe strong fluctuations and something like waves in the distribution of publications per model and in keywords. 
We do not have enough data to verify this hypothesis. However, we see that the most common rated power we find in papers with our search methods is 2~MW, while a few papers discuss turbines with other rated power in the range from 600~kW to 4~MW. 
The notable effect is that the keywords vibration, torque and strain allow us to find papers talking about structural elements such as the tower and foundation of large offshore wind turbines with the rated power of the order of 10~MW~\cite{de2023long}.

\section{Conclusion}\label{Sec:Conclusions}
We identified articles that make long-term forecasts of wind turbine failures and classified these articles based on the prediction time, the size of the dataset, and the methods used. 
The requirement of predicting failures at least 2 days in advance appears to be a filter that selects datasets larger than 0,4~TY. 
A transition from \texttt{small} to \texttt{large} datasets is observed at about 10~TY. 
Below this threshold, authors tend to use specific methods and advanced modeling, while after that more machine learning and statistical approaches with filtering are used. 
Additionally, for datasets larger than 10~TY, the forecasts are made for about two weeks, which can be taken as a data-driven definition of a \texttt{long}-term forecast. 
An approximate linear scaling \texttt{prediction days}\,$\sim~2*$TY is found.
Vibrational data are used more often for small datasets, giving an order of magnitude better prediction horizons. However, the use of specific advanced modeling with the standard SCADA data gives comparable results. 
We do not see any grouping by failed components. 
The number of methods used by the authors of the selected articles exceeds the number of articles, while the methods rarely repeat. 
This means that more analysis attempts and dedicated comparisons of different methods on the same datasets are needed from the research community to compare the performance of different methods.


\appendix

\begin{acronym}[DI-CUSUM]

\acro{ANFIS}{adaptive neuro-fuzzy interference system}
\acro{AIn}{anomaly index}
\acro{AOIn}{anomaly operation index}
\acro{AR}{auto regression}
\acro{AE}{autoencoder}
\acro{BagR}{bagging regression}
\acro{BayC}{Bayesian classifier}
\acro{BayR}{Bayesian regularization}
\acro{bi-LSTM}{bidirectional LSTM}
\acro{BGraph}{Bond Graph}
\acro{CombP}{combination prediction}
\acro{CNN}{convolutional neural network}
\acro{CUSUM}{cumulative sum}
\acro{DI-CUSUM}{decision interval CUSUM}
\acro{DT}{decision tree}
\acro{dAE}{deep autoencoder}
\acro{DBN}{deep belief network}
\acro{DDPG-SOM}{deep deterministic policy gradient-self-organizing map}
\acro{DAE}{denoising auto-encoder}
\acro{DBSCAN}{density-based spatial clustering of applications with noise}
\acro{DegrV}{degradation velocity}
\acro{EWMA}{exponentially weighted moving average}
\acro{FI}{feature importance}
\acro{FV}{feature vector}
\acro{FL}{fuzzy logic}
\acro{GRU}{gated recurrent unit}
\acro{GMC}{Gaussian mixture copula model}
\acro{GMIn}{global Mahalanobis indicator}
\acro{GBDT}{gradient boosting DT}
\acro{GCA-MTS}{graph convolutional autoencoder for multivariate time series}
\acro{HIn}{health index}
\acro{HCF}{high correlation filter}
\acro{ICA}{independent component analysis}
\acro{IE}{information entropy}
\acro{IF}{isolation forest}
\acro{kNN}{k nearest neighbours}
\acro{KF}{Kalman filter}
\acro{KCPD}{kernel-based change-point detection}
\acro{kMC}{k-means clustering}
\acro{LASSO}{least absolute shrinkage and selection operator}
\acro{LARS}{least-angle regression}
\acro{LinR}{linear regression}
\acro{L-SVC}{linear support vector machine classifier}
\acro{LoMST}{local minimum spanning tree}
\acro{LOF}{local outlier factor}
\acro{LRIn}{local residual indicator}
\acro{LogR}{logistic regression}
\acro{LSTM}{long short-term memory neural network}
\acro{MD}{Mahalanobis distance}
\acro{MA}{moving average}
\acro{MLR}{multivariate linear regression}
\acro{MR}{multivariate regression}
\acro{MI}{mutual information}
\acro{NN}{neural network without a specific name}
\acro{NARX}{nonlinear autoregressive NN with exogenous inputs}
\acro{PF}{particle filtering}
\acro{PPMC}{Pearson prod-moment correlation}
\acro{Phys}{physics based model}
\acro{PPS}{power predictive score}
\acro{PCA}{principal component analysis}
\acro{QF}{quantile filtering}
\acro{RBFNN}{radial basis function neural network}
\acro{RF}{random forest}
\acro{RNN}{recurrent neural network}
\acro{RBM}{restricted Boltzmann machine}
\acro{SVM}{support vector machine}
\acro{SVR}{support vector regression}
\acro{SMOTE}{synthetic minority over-sampling technique}
\acro{WHC}{Ward hierarchical clustering}
\acro{WT}{wavelet transform}
\acro{WD}{Weibull distribution}
\acro{WA}{weighted average}
\acro{WP}{Wiener process}
\acro{XGBOOST}{extreme gradient boosting}

\end{acronym}

\section*{Acknowledgment}

We thank Sergiy Begun for fruitful discussions. 
%
This research was funded by the Bavarian State Ministry of Science and the Arts, Germany.


%
%

\bibliography{Begun_Long-Term_Forecast_IEEE}

%

%

\begin{IEEEbiography}[{\includegraphics[width=1in,height=1.25in,clip,keepaspectratio]{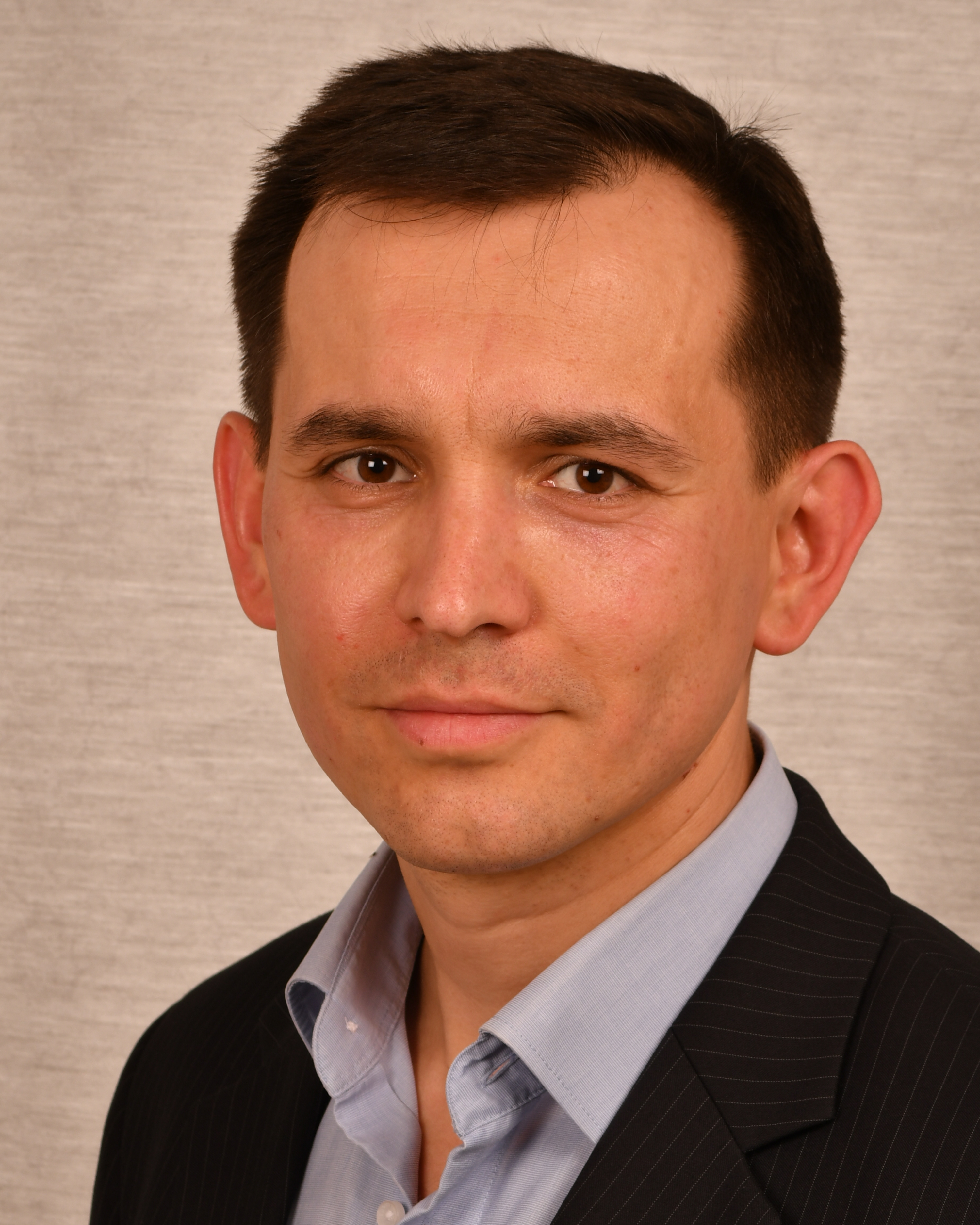}}]{Viktor Begun} studied physics at Taras Shevchenko National University and obtained his PhD in Theoretical Physics from Bogolyubov Institute for Theoretical Physics. He worked as a Postdoc at Goethe University Frankfurt and at Jan Kochanowski University, as an Assistant Professor at Warsaw University of Technology, as a Lecturer at the Augsburg University of Applied Sciences, and as a Researcher at the Technische Hochschule Ingolstadt.  
\end{IEEEbiography}
\begin{IEEEbiography}[{\includegraphics[width=1in,height=1.25in,clip,keepaspectratio]{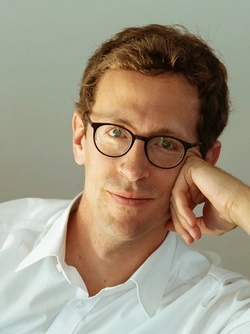}}]{Ulrich Schlickewei} is Professor of Mathematics and Data Science at Technische Hochschule Ingolstadt. He studied Mathematics at the Universities of Freiburg, Rome, and Paris and obtained his PhD in the field of Algebraic Geometry at the University of Bonn. As a postdoctoral researcher, he was part of the Computer Vision lab headed by Daniel Cremers at the Technical University of Munich. Ulrich served as a project manager at McKinsey \& Company, as Head of Strategy at appliedAI, and held professorships at the University San Francisco de Quito and Technical University of Applied Sciences Augsburg.

\end{IEEEbiography}

\end{document}